\documentclass[
showkeys,
preprintnumbers,amsmath,amssymb,APS]{revtex4}

\usepackage{graphicx}
\usepackage{dcolumn}
\usepackage{bm}

\usepackage{graphicx}
\usepackage{epsfig}
\usepackage{color}

\begin{document}

\title{Barotropic FRW cosmologies with Chiellini damping in comoving time}

\author{Haret C. Rosu}
\email{hcr@ipicyt.edu.mx}
\affiliation{IPICyT, Instituto Potosino de Investigacion Cientifica y Tecnologica,\\
Camino a la presa San Jos\'e 2055, Col. Lomas 4a Secci\'on, 78216 San Luis Potos\'{\i}, S.L.P., Mexico}

\author{Stefan C. Mancas }
\email{stefan.mancas@erau.edu}
\affiliation{Department of Mathematics, Embry-Riddle Aeronautical University,\\ Daytona Beach, FL. 32114-3900, U.S.A.}

\author{Pisin Chen}
\email{pisinchen@phys.ntu.edu.tw}
\affiliation{Leung Center for Cosmology and Particle Astrophysics (LeCosPA) and Department of Physics, National Taiwan University, Taipei 10617, Taiwan\\}

\begin{abstract}
For non-zero cosmological constant $\Lambda$, we show that the barotropic FRW cosmologies as worked out in the comoving time lead in the radiation-dominated case to scale factors of identical form as for the Chiellini dissipative scale factors in conformal time obtained recently by us in Phys. Lett. A 379 (2015) 882-887. This is due to the Ermakov equation which is obtained in this case. For zero cosmological constant, several textbook solutions are provided as particular cases of $\Lambda\neq 0$.
\end{abstract}


\maketitle

{\bf 1. Parametric solutions of the barotropic FRW equations for non-zero cosmological constant}\\

Barotropic fluids continue to be of interest at the frontier of cosmology. In the past few years, they have been used as illustrative minimal models in the challenging problem of the dark sector of the Universe \cite{0}, and currently they are considered as a viable partial solution for the late acceleration of the Universe due to their rapid fading to a de Sitter phase \cite{00}.

In this Letter, we are concerned with the solutions of the Einstein-Friedmann dynamical equations of barotropic FRW cosmologies with a cosmological constant $\Lambda $ for the scale factor of the universe, $a(t)$, with $\dot \,=d/dt$

\begin{eqnarray}\label{eq1}
\begin{array}{lll}
\frac{\ddot{a}}{a}&=-\frac{4\pi G}{3}\left(\rho+3p\right)+\frac{\Lambda}{3}\\
H^2&=\left(\frac{\dot{a}}{a}\right)^2=\frac{8\pi G \rho}{3}-\frac{\kappa}{a^2} +\frac{\Lambda}{3}~,
\end{array}
\end{eqnarray}
where $\kappa$ is the curvature of the universe, which is $0, \pm1$, if the universe is flat, closed or open, respectively,  $\rho, \,p$ are the energy density and pressure, $t$ is the comoving time, $G$ is the universal gravitational constant, $\Lambda$ is Einstein\rq{}s cosmological constant which is related to the pressure of the vacuum, and $H$ is the Hubble  expansion parameter. Since for a barotropic fluid the equation of state is
\begin{equation}\label{eq2}
p=(\gamma-1)\rho,
\end{equation}
then the FRW equations can be decoupled and yield to
\begin{equation}\label{eq3}
a \ddot a+\bar{\gamma}\dot a ^2+ \bar \gamma \kappa=\frac{\Lambda}{3} (\bar \gamma+1)a^2,
\end{equation}
where $\bar \gamma=\frac 3 2 \gamma-1 \ne 0$.

Let us introduce the transformation 
\begin{equation}\label{eq5}
\dot a^2=z\left(a(t)\right)
\end{equation}
then $z$ satisfies the linear equation
\begin{equation}\label{eq6}
\frac{dz}{da}+\frac{2\bar \gamma}{a}z=\frac {2\Lambda}{ 3}(\bar \gamma+1)a-\frac{2 \bar \gamma \kappa}{a}~.
\end{equation}
By one quadrature with positive initial conditions we have
\begin{eqnarray}\label{eq7}
\begin{array}{lll}
z(a)&=C_1a^{-2 \bar \gamma}-\kappa+\frac{\Lambda}{3}a^2, \qquad &  \mathrm{if} \quad \bar{\gamma}\ne -1\\
z(a)&=C_2a^2-\kappa, \qquad & \mathrm{if} \quad  \bar{\gamma}= -1\\
\end{array}
\end{eqnarray}
and by using equation \eqref{eq5} we obtain the following parametric solutions
\begin{eqnarray}\label{eq9}
\begin{array}{lll}
 {\displaystyle t-t_0=\int {\frac{da}{\sqrt{C_1a^{-2 \bar \gamma}-\kappa+\frac{\Lambda}{3}a^2}}}}, \qquad &  \mathrm{if} \quad \bar{\gamma}\ne -1\\
{\displaystyle t-t_0=\int {\frac{da}{\sqrt{C_2a^2-\kappa}}}}, \qquad & \mathrm{if} \quad  \bar{\gamma}= -1~.\\
\end{array}
\end{eqnarray}

In addition, since $z(a)$ is known from \eqref{eq7}, we substitute it back into \eqref{eq3}, and using \eqref{eq5} we obtain the equation of motion for the scale parameter in which the curvature of the universe does not occur explicitly
\begin{eqnarray}\label{eq9a}
\begin{array}{lll}
\ddot a-\frac{\Lambda}{3}a=-C_1\bar \gamma a^{-(2\bar\gamma+1)}, \qquad &  \mathrm{if} \quad  \bar{\gamma}\ne -1\\
\ddot a-C_2a=0, \qquad & \mathrm{if} \quad  \bar{\gamma}= -1~.\\
\end{array}
\end{eqnarray}

{\bf 2. The radiation-dominated case, $\bar{\gamma}=1$}\\

Interestingly enough, for the universe dominated by radiation one may recognize \eqref{eq9a} as the following Ermakov's equation
\begin{equation}\label{eq9b}
\ddot a-\frac{\Lambda}{3}a=-C_1a^{-3}~.
\end{equation}
From this equation, by multiplication with $\dot a$ and one integration, one can obtain 
\begin{equation}\label{eq9c}
 {\displaystyle t-t_0=\int \frac{a \, da}{\sqrt{\frac \Lambda 3  a^4-\kappa a^2+C_1}}~,}
 \end{equation}
which is the solution given in the first line of \eqref{eq9} in the case of radiation.\\

$\kappa =\pm 1$. The solutions of the above integral depend on the sign of the discriminant $\Delta\equiv \kappa^2-\frac{4C_1\Lambda}{3}=1-\frac{4C_1\Lambda}{3}$, and are shown in Table \ref{Tab3}. 

\begin{table}[ht!]
\begin{center}
\begin{tabular}{|c|c|c|c|}
\hline  & $\kappa=-1$ (top signs and pair of signs in the main square root)\\ \hline & $\quad \kappa=1$ (bottom signs and pair of signs in the main square root) \\
\hline
\hline
$\bar\gamma=1$  &   $a(t) =\pm  \left\{
     \begin{array}{lr}
         \frac 1 \lambda\sqrt{\mp 1\, {}_{\mp}^{\pm}\sqrt {-\Delta}\, \mathrm{sinh}\sqrt{2}\lambda(t-t_0)} & : \Lambda>\frac{3}{4C_1}\\
        \frac{1}{\lambda}\sqrt{\mp 1\pm\lambda^2e^{\pm \sqrt{2} \lambda(t-t_0)}} & : \Lambda=\frac{3}{4C_1}\\
         \frac 1 \lambda  \sqrt{\mp 1\pm \sqrt {\Delta}\, \mathrm{cosh}\sqrt{2}\lambda(t-t_0)} &:0< \Lambda<\frac{3}{4C_1}\\
        \frac 1 \lambda\sqrt{\pm 1{}_{\pm}^{\mp}\sqrt {\Delta}\, \mathrm{sin}\sqrt{2} \lambda(t-t_0)}&: \Lambda<0<\frac{3}{4C_1}\\
     \end{array}
   \right.$ \rule[0.5cm]{0cm}{0.8cm}\\
\hline
\end{tabular}
\end{center}
\caption{Solutions for open and closed radiation-dominated universes, $\Lambda \ne 0$, $\lambda=\sqrt{\frac{|\Lambda|}{3}}$.}
\label{Tab3}
\end{table}

We notice that these solutions are similar to the recently introduced Chiellini dissipative scaling factors of barotropic FRW universes in conformal time because if we identify $-2\kappa\tilde{\gamma}^2\equiv\lambda^2$, $c_1\equiv -\kappa$, and $2{\rm k}\equiv C_1$ in the Chiellini-damped solutions given in \cite{pla} 
their discriminants 
turns into the discriminant in the non-zero cosmological constant cases discussed here. These analytical solutions are based on the Chiellini integrability condition for Abel equations of the first kind \cite{chiell}, which also describe the dynamics of FRW cosmologies filled with bulk viscous fluids \cite{mak1, mak2}, as well as FRW universes filled with scalar fields \cite{yy}.\\

$\kappa=0$. In this case, we have
\begin{equation}\label{eq9c}
 {\displaystyle t-t_0=\int \frac{a \, da}{\sqrt{\frac \Lambda 3  a^4+C_1}}.}
 \end{equation}
 The solutions of the above integral are given in Table \ref{Tab7}.

\begin{table}[ht!]
\begin{center}
\begin{tabular}{|c|c|c|c|}
\hline  & $\kappa=0$  \\
\hline
\hline
$\bar\gamma=1$  &   $a(t) =\pm \frac{\sqrt[4]{C_1}}{\sqrt \lambda} \left\{
     \begin{array}{lr}
        \sqrt{\mathrm{sinh}\, 2\lambda(t-t_0)} & : \Lambda>0\\
        \sqrt{\mathrm{sin}\,2\, \lambda(t-t_0)} & : \Lambda<0\\
     \end{array}
   \right.$ \rule[0.5cm]{0cm}{0.2cm}\\
\hline
\end{tabular}
\end{center}
\caption{Solutions for the flat radiation-dominated universe, $\Lambda \ne 0$, $\lambda=\sqrt{\frac{|\Lambda|}{3}}$.}
\label{Tab7}
\end{table}

{\bf 3. Cases of zero cosmological constant, $\Lambda=0$}\\

Zero cosmological constant cases can be treated as particular cases of the formulas \eqref{eq9}. Let us denote the two integration constants from \eqref{eq9} as $C_1=a_0^{2 \bar \gamma}$, $C_2=a_0^{-2}$, which mean that all solutions pass through $a(t_0)=a_0$, then the solutions of \eqref{eq9} are
\begin{eqnarray}\label{eq10}
\begin{array}{lll}
a(t)&=a_0[1+(\bar \gamma+1)\frac{t-t_0}{a_0}]^{\frac{1}{\bar \gamma+1}}, \qquad &  \mathrm{if} \quad \bar{\gamma}\ne -1, \,  \kappa=0\\
a(t)&=a_0e^{\frac{t-t_0}{a_0}}, \qquad & \mathrm{if} \quad \bar{\gamma}= -1, \,  \kappa=0\\
\end{array}
\end{eqnarray}

The first two solutions are obtained easily by integrating equations \eqref{eq9}, and they represent the restricted form of flat solution  when $t_0=\frac{a_0}{\bar \gamma +1}$ for which $a(t)=a_0 \big(\frac{t}{t_0}\big)^{\frac{1}{\bar \gamma+1}}$, in the first case, while for $t_0=a_0 \ln a_0$, we obtain the de Sitter flat solution $a(t)=e^{\frac { t} {t_0}}$ in the second case.

For the three textbook cases of $\bar \gamma =1 $ (radiation), $\bar \gamma =\frac 1 2$ (dust), and $\bar \gamma =-1$ (vacuum), by solving \eqref{eq9a} with $\Lambda =0$ the solutions are presented in Tables \ref{Tab1} and  \ref{Tab2}.

\begin{table}[ht!]
\begin{center}
\begin{tabular}{|c|c|c|c|} 
\hline
$\bar\gamma=1$  & $a(t)=a_0\sqrt{-1+(1+\frac{t-t_0}{a_0})^2}$ \rule[-0.3cm]{0cm}{0.8cm}\\
\hline
$\bar\gamma=\frac 1 2$ & $\sqrt{A(t)}\sqrt{1+A(t)}-\mathrm{arccosh}\sqrt{1+A(t)}=\frac{t-t_0}{a_0}+\sqrt 2-\mathrm{arccosh}\sqrt 2$ \rule[-0.3cm]{0cm}{0.8cm}\\
\hline
$\bar\gamma=-1$   & $a(t)=a_0\sinh\left(\frac{t-t_0}{a_0}+\mathrm{arcsinh}1\right)$  \rule[-0.3cm]{0cm}{0.8cm}\\
\hline
\end{tabular}
\end{center}
\caption{Solutions for $\kappa=-1$ , $\Lambda = 0$, $A(t)=\frac{a(t)}{a_0}$.}
\label{Tab1}
\end{table}

\begin{table}[ht!]
\begin{center}
\begin{tabular}{|c|c|c|c|}
\hline
$\bar\gamma=1$  & $a(t)=a_0\sqrt{1-(1-\frac{t-t_0}{a_0})^2}$ \rule[-0.3cm]{0cm}{0.8cm} \\
\hline
$\bar\gamma=\frac 1 2$ & $\sqrt{A(t)}\sqrt{1-A(t)}+\mathrm{arcsin}\sqrt{1-A(t)}=\frac{t-t_0}{a_0}$ \rule[-0.3cm]{0cm}{0.8cm} \\
\hline
$\bar\gamma=-1$   & $a(t)=a_0\cosh\left(\frac{t-t_0}{a_0}\right)$ \rule[-0.3cm]{0cm}{0.8cm} \\
\hline
\end{tabular}
\end{center}
\caption{Solutions for $\kappa=1$ , $\Lambda = 0$,  $A(t)=\frac{a(t)}{a_0}$.}
\label{Tab2}
\end{table}

In the last case, $\bar{\gamma} \ne 0, \,  \kappa\ne 0 $,  we will follow the substitution given by Lima \cite{Lima}, see also \cite{nor}
\begin{equation}\label{eqma1}
u(a)= \kappa A^{2 \bar \gamma}~, 
\end{equation}
where $A(t)=\frac{a (t)}{a_0}$,
and differentiating $u(a)$ with respect to $t$ we have
\begin{equation}\label{eqma2}
\dot u= 2 \bar \gamma u \frac {\dot a}{a}.
\end{equation}
By inverting equation~\eqref{eqma1}  which will give  $a$  and finding  $\dot a$ from \eqref{eq5}, we obtain the first order equation that gives $t(u)$ as the incomplete beta function
\begin{equation}\label{eqma3}
\frac{dt}{du}=Du^{\mu-1}(1-u)^{- \frac 1 2},
\end{equation}
where $\mu=(\bar \gamma+1)/2 \bar \gamma$ and $D=a_0\kappa^{1/2\bar \gamma}/2 \bar \gamma$. We differentiate again with respect to $u$ and obtain the equation
\begin{equation}\label{eqma4}
u(1-u)\frac{d^2t}{du^2}+\left[(1-\mu)-\left(\frac 3 2 -\mu\right)u\right]\frac{dt}{du}=0~,
\end{equation}
which is the hypergeometric equation
with  coefficients $a=0$, $b=\frac 1 2 -\mu=-\frac{1}{2 \bar \gamma}$, and $c=1-\mu=\frac 12 -\frac{1}{2 \bar \gamma}$.

The general solution of \eqref{eqma4} is the linear combination
\begin{equation}\label{eqma6b}
t(u)=\alpha \, {}_2F_1\left(0,-\frac{1}{2 \bar \gamma};\frac 12-\frac{1}{2 \bar \gamma};u\right)+\beta
u^{\frac 1 2+\frac{1}{2 \bar \gamma}}\, {}_2F_1\left (\frac 1 2,\frac 1 2+\frac{1}{2 \bar \gamma};\frac 32+\frac{1}{2 \bar \gamma};u\right)~.
\end{equation}
However, according to formula 15.4.1 in \cite{AS}, the first hypergeometric function is equal to unity. Thus, the general solution can be written in the form
\begin{equation}\label{eqma6a}
t(u)=\alpha+\beta u^{\frac 1 2+\frac{1}{2 \bar \gamma}}\, {}_2F_1\left (\frac 1 2,\frac 1 2+\frac{1}{2 \bar \gamma};\frac 32+\frac{1}{2 \bar \gamma};u\right)~.
\end{equation}
It is convenient to pass to the complementary variable $1-u$ in the above hypergeometric function using the formula 15.3.6 in \cite{AS}.
Then, one obtains
\begin{equation}\label{eqma6c}
t(u)=\alpha+\beta \sqrt{\pi}\left(\frac 1 2+\frac{1}{2 \bar \gamma}\right) u^{\frac 1 2+\frac{1}{2 \bar \gamma}}\, {}_2F_1\left(\frac 1 2+\frac{1}{2 \bar \gamma},\frac 12;\frac 12;1-u\right)-2\beta (1-u)^{\frac 12} u^{\frac 1 2+\frac{1}{2 \bar \gamma}}\, {}_2F_1\left (1,\frac 1 2+\frac{1}{2 \bar \gamma};\frac 32;1-u\right)~.
\end{equation}
The first hypergeometric function in \eqref{eqma6c} can be shown to be $u^{-(\frac 1 2+\frac{1}{2 \bar \gamma})}$, which leads to the final result
\begin{equation}\label{eqma6}
t(u)=t_0-2\beta (1-u)^{\frac 12} u^{\frac 1 2+\frac{1}{2 \bar \gamma}}\, {}_2F_1\left (1,\frac 1 2+\frac{1}{2 \bar \gamma};\frac 32;1-u\right)~,
\end{equation}
where $t_0$ (the origin of time) depends on $\bar \gamma$
\begin{equation}\label{eqma6-1}
t_0=\alpha+\beta \sqrt{\pi}\left(\frac 1 2+\frac{1}{2 \bar \gamma}\right)~.
\end{equation}
By choosing carefully the integration constants $\alpha$ and $\beta$ in \eqref{eqma6}, one can obtain the textbook solutions presented in Tables \ref{Tab1} and \ref{Tab2}.

The general solution \eqref{eqma6} is not valid when $c$ is either equal to unity or any negative integer, i.e., when $\bar \gamma = 1/(2n+1)$, $n=-1,1,2...$~. In these cases, the general solution has been given in \cite{nor}, see also \cite{Lima}, and involves the logarithmic solutions of the hypergeometric equation as the other linearly independent solution. This is due to degeneracy for $c=1$ and meaningless of one of the hypergeometric series in all the other negative integer cases of $c$.

\bigskip
\bigskip

{\bf Acknowledgments}\\

S.M. would like to acknowledge partial support from the Dean of Research \& Graduate Studies at Embry-Riddle Aeronautical University. P.C. appreciates the supports from the Ministry of Science and Technology (MOST), Taiwan, under Grant 101-2923-M-002-006-MY3.

\end{document}